# Fluorescence Resonance Energy Transfer (FRET) sensor

Syed Arshad Hussain*, Dibyendu Dey, Sekhar Chakraborty, Jaba Saha, Arpan Datta Roy, Santanu Chakraborty, Pintu Debnath, D. Bhattacharjee

Department of Physics, Tripura University, Suryamaninagar – 799022, Tripura, India
* Corresponding author
Email: sa_h153@hotmail.com, sahussain@tripurauniv.in
Ph: +919862804849 (M), +91381 2375317 (O)
Fax: +913812374802

**Abstract**

The applications of Fluorescence resonance energy transfer (FRET) have expanded tremendously in the last 25 years, and the technique has become a staple technique in many biological and biophysical fields. FRET can be used as spectroscopic ruler in various areas such as structural elucidation of biological molecules and their interactions, in vitro assays, in vivo monitoring in cellular research, nucleic acid analysis, signal transduction, light harvesting, and metallic nanomaterials etc. Based on the mechanism of FRET a variety of novel chemical sensors and Biosensors have been developed. This review highlights the recent applications of sensitive and selective ratiometric FRET based sensors.

**Keywords**
FRET, ratiometric sensor , permanent hardness , laponite, pH sensor, ion sensor, biosensor.

**1. Introduction**

The Fluorescence resonance energy transfer (FRET) has been widely used as a spectroscopic technique in all applications of fluorescence, including medical diagnostics, DNA analysis, optical imaging [1] and for various sensing properties [2-11]. Generally, fluorescence-based sensors adopt three different strategies: (a) fluorescence quenching (turn-off), (b) fluorescence enhancement (turn-on), and (c) fluorescence resonance energy transfer (FRET). FRET sensors became popular tools for studying intracellular processes [12-16]. FRET between two molecules is an important physical phenomenon with considerable interest for the understanding of some biological systems and with potential applications in optoelectronic and thin film device development [17, 18].

In this account, we review the application of FRET for designing various sensors, focusing primarily on ion sensor, bio-sensor, hard water sensor and pH sensor. The technique of FRET, when applied to optical microscopy, permits to determine the approach between two molecules within several nanometers. FRET was first described over 50 years ago, that is being used more and more in biomedical research and drug discovery today. FRET is an electrodynamic phenomenon that occurs through non-radiative process whereby an excited state donor D (usually a fluorophore) transfers energy to a proximal ground state acceptor A through long-range dipole−dipole interactions [19]. The acceptor must absorb energy at the emission wavelength(s) of the donor but does not necessarily have to remit the energy fluorescently itself. The rate of energy transfer depends on a number of factors including the fluorescence quantum yield of the donor in the absence of acceptor, the refractive index of the solution, the dipole angular orientation of each molecule. FRET can be an accurate measurement of molecular proximity at angstrom distances (10-100Å). Due to its sensitivity to distance, FRET has been used to investigate molecular level interactions [20-27]. Though the physics and chemistry behind the fluorescence resonance energy transfer have been well studied theoretically for years but only with recent technical advances has it become feasible to apply FRET to sensing research [2-11]. Recent advances in the technique have led to qualitative and quantitative improvements, including increased spatial resolution, distance range and sensitivity. FRET mechanisms are also important to other phenomena, such as photosynthesis kinetics, chemical reactions and Brownian dynamics [28, 29]. Recently, FRET phenomenon have been employed for the conformation of proteins and knowing their structure [30], for the detection of spatial distribution and assembly of proteins [31], for the designing biosensor [10], for nucleic acid hybridization [32], distribution and transport of lipids [33].

One of the important methods for sensing of different chemical and biological materials is fluorescent sensors. In recent years, fluorescence spectroscopy has become a powerful tool for the detection of transition and heavy metal ions with high sensitivity and simplicity [34-37]. However, many of the chemosensors have only one signal for detecting, i.e., the fluorescence intensity, and could be easily perturbed by the environmental and instrumental conditions [38-39]. However, it is interesting to mention that the introduction of ratiometric chemosensors can eliminate or reduce the effects of these factors by the self-calibration of the two emission bands [40-51]. Recently, the FRET has been employed to design ratiometric sensors [50, 49, 52-56]. The design of ratiometric sensors can be done by two method (i) ICT (intermolecular charge transfer) and (ii) FRET (Fluorescence Resonance Energy Transfer). For many ICT based ion sensors it is difficult to determine the ratio between two relatively broad signal emissions. The advantage of FRET over ICT is that the ratio between two fluorescence intensities is independent of these external factors such as fluctuation of excitation source and sensor concentration. FRET observed the changes in the intensity ratio of absorption and emission which is favourable in increasing the signal selectivity. A significant advantage of FRET-based sensing is that it simplifies the design of the fluorophore. Recently, FRET based sensing has become most effective method for the detection of ions in environment. FRET based sensors have been widely used in metal ion detection [57, 58], sensing of the fluorophores [59–63], Silica [64, 65], and polymer particles [66–68].

In this review, we focus on discussion of various types of chemical and biological sensors like pH sensor, hard water sensor, Ion sensor and DNA sensor, which were developed by the implementation of FRET phenomenon between two laser dyes Acriflavine (Acf) and Rhodamine B (RhB) mainly by our group. As both the dyes Acf & RhB are highly fluorescent and the fluorescence spectrum of Acf sufficiently overlaps with the absorption spectrum of RhB, they are in principle suitable for FRET. These phenomenons have been investigated in aqueous solution, clay dispersion and layer-by-layer (LbL) self assembled films. The effect of various types of ions (like $K^+$, $Na^+$, $Mg^{+2}$, $Ca^{+2}$, $Fe^{+2}$, $Fe^{+3}$, $Al^{+3}$ etc) on the FRET efficiency between Acf & RhB have been discussed in Ion Sensor where as in pH sensor the effect of different pH has been investigated. Hard water sensor is able to sense the extent of hardness of water and DNA sensor can be very useful to detect the effect of DNA on the spectrum of Acf & RhB. Of late the sensing technology has been enormously enhanced due to suitable application of FRET.

**2. Principle and theoretical consideration of FRET**

In the process of FRET, initially a donor fluorophore (D) absorbs the energy due to the excitation of incident light and transfer the excitation energy to a nearby chromophore, the acceptor (A).

$$D + h\upsilon \rightarrow D^*$$

$$D^* + A \rightarrow D + A^*$$

$$A^* \rightarrow A + h\upsilon$$

[Where $h$ is the Planck's constant and $\upsilon$ is the frequency of the radiation]

Energy transfer manifests itself through decrease or quenching of the donor fluorescence and a reduction of excited state lifetime accompanied also by an increase in acceptor fluorescence intensity. Figure 1 is a Jablonski diagram that illustrates the coupled transitions involved between the donor emission and acceptor absorbance in FRET. In presence of suitable acceptor, the donor fluorophore can transfer its excited state energy directly to the acceptor without emitting a photon.

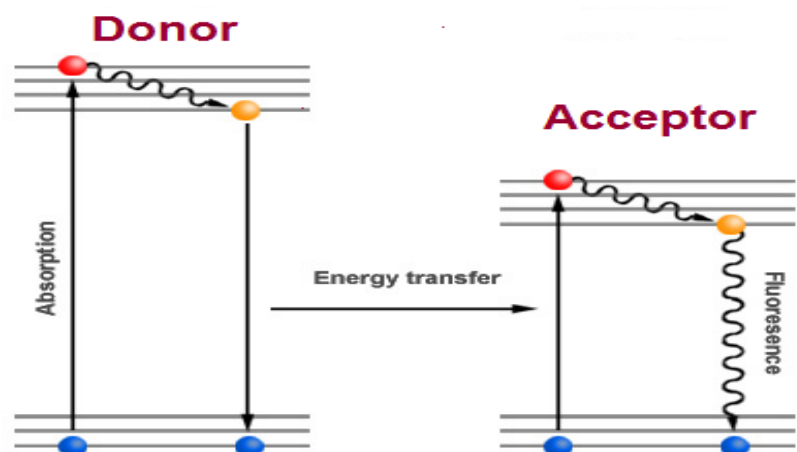

**Figure 1.** Jablonski diagram illustrating the FRET process.

There are few criteria that must be satisfied in order for FRET to occur. These are: (i) the fluorescence emission spectrum of the donor molecule must overlap the absorption or excitation spectrum of the acceptor chromophore. The degree of overlap is referred to as spectral overlap integral (J). (ii) The two fluorophore (donor and acceptor) must be in the close proximity to one another (typically 1 to 10 nanometer). (iii) The transition dipole orientations of the donor and acceptor must be approximately parallel to each other. (iv) The fluorescence lifetime of the donor molecule must be of sufficient duration to allow the FRET to occur.

Solving the enigma surrounding fluorescence quenching experiments revealed the phenomenon of FRET and led J. Perrin [69] to propose dipole–dipole interactions as the mechanism via which molecules can interact without collisions at distances greater than their molecular diameters. Some 20 years later, Förster [70, 71] built upon Perrin's idea to put forward an elegant theory which provided a quantitative explanation for the non-radiative energy transfer in terms of his famous expression given by

$$k_T(r) = \frac{1}{\tau_D}\left(\frac{R_0}{r}\right)^6 \rightarrow (1)$$

Where, $k_T(r)$ is the rate of energy transfer from donor to acceptor, r is the distance between donor and acceptor and $R_0$ is the well-known Förster radius given by the spectral overlap between the fluorescence spectrum of the donor and the absorption spectrum of the acceptor. The distance at which resonance energy transfer is 50% efficient, is called the Förster distance. At r = $R_0$, the transfer efficiency is 50% and at this distance the donor emission would be decreased to half of its intensity in the absence of acceptor.
The value of $R_0$ can be defined by the following expression [72-75]

$$R_0^6 = \left[\frac{9000(\ln 10)k^2\phi_D}{128\pi^5 N n^4}\right]\int_0^\alpha F_D(\lambda)\varepsilon_A(\lambda)\lambda^4 d\lambda \rightarrow (2)$$

Where,
$F_D$ = the normalized fluorescence intensity of the donor.
$\varepsilon_A(\lambda)$ = the extinction coefficient of the acceptor (in $M^{-1}cm^{-1}$).
$\lambda$ = the wavelength (in nm).
$\phi_D$ = the fluorescence quantum yield of the donor in the absence of acceptor.
$n$ = is the refractive index of the medium.
$k^2$ = orientation factor of transition dipole moment between donor (D) and acceptor (A).
N = Avogadro number.
The integral part of equation (2) is known as the spectral overlap integral $J(\lambda)$ and is given by

$$J(\lambda) = \int_0^\alpha F_D(\lambda)\varepsilon_A(\lambda)\lambda^4 d\lambda \rightarrow (3)$$

Therefore the above definition of $R_0$ in equation (2) can be rewritten in terms of $J(\lambda)$ with units $M^{-1}cm^{-1}nm^4$ as

$$R_0 = 0.2108\{k^2 n^{-4}\Phi_D J(\lambda)\}^{1/6} \rightarrow (4)$$

Where $R_0$ is in units of $A^0$.
The energy transfer efficiency can be termed as [72-74]

$$E = \frac{k_T(r)}{k_T(r)+\tau_D^{-1}} = \frac{\tau_D k_T(r)}{1+\tau_D k_T(r)} \rightarrow (5)$$

This is the fraction of the transfer rate to the total decay rate of the donor. Using equation (1) and (5) E can be expressed as

$$E = \frac{R_0^6}{R_0^6 + r^6} \rightarrow (6)$$

The efficiency of the energy transfer (E) can also be expressed as [57]

$$E = 1 - \frac{F_{DA}}{F_D} \rightarrow (7)$$

where $F_{DA}$ is the relative fluorescence intensity of donor in the presence of acceptor and $F_D$ is the fluorescence intensity of donor in the absence of acceptor. This equation is equivalent to equation 5 [75].

**3. FRET as molecular spy**

The unique feature of FRET is its capability to inform us whenever two molecules (donor and acceptor) are close to one another on a molecular scale (usually within 1-10 nm), and whether they are moving relative to each other. It is also possible to detect how the donor and acceptor transition moments are oriented relative to each other. This is because the FRET efficiency depends on donor-acceptor distance as well as on the relative orientations of the two dipoles. It is possible to couple FRET pair with other physical and biological methods, and this greatly extends the usefulness of the process. Therefore fluropore involving FRET can be considered as analogous to roaming molecular spies with radio transmitters, radiating information to the experimenter about the state of affairs on the molecular scale, and informing us where the spies are located and how they are oriented.

Generally FRET (spectroscopic experiments) can be carried out in most laboratories, whether the ''samples'' are large (such as in cuvettes, or even on whole mammalian bodies) or small (such as in the fluorescence microscope, and on the single molecule level). Therefore, irrespective of the scale of the sample, the information on the molecular scale derivable from FRET remains accessible. Accordingly, FRET can be considered as like a spectroscopic microscope, providing us information about the distance and orientation on the molecular scale regardless of the size of the sample. Also it is possible to follow the dynamics of changes in molecular dimensions and proximities by monitoring FRET with time.

## 4. Principle of FRET sensor design

Typical FRET sensor consists of a recognition element (sensing material) fused to a pair of fluorophores (FRET pair) capable of FRET or a system containing the FRET pair and the recognition element. A conformational change in the recognition element can be exploited to bring about changes in FRET efficiency when fused to an appropriate FRET pair. Or presence of recognition element affects FRET efficiency. Also analyte dependent changes in the spectra of FRET pair can change the FRET efficiency [71-73]. Therefore, FRET is a unique phenomenon in generating fluorescence signals sensitive to molecular conformation, association and separation in the 1–10 nm range [72].

FRET is a non-radiative quantum mechanical process where energy transfer occurs between two fluorophores in close proximity (less than 10 nm apart) through long-range dipole−dipole interactions [74, 75]. The rate of energy transfer is highly dependent on many factors, such as the extent of spectral overlap, the relative orientation of the transition dipoles, and, most importantly, the distance between the donor and acceptor molecules [76, 77]. FRET usually occurs over distances comparable to the dimensions of most biological macromolecules, that is, about 10 to 100 Å. This makes FRET a spectroscopic ruler to study biological systems [78-80]. Since FRET is extremely sensitive to changes in the relative positions of two fluorophores or their orientation, even a subtle change in the conformation of the recognition element, when attached to a FRET pair, can be visualized as a change in FRET efficiency. Also presence of minute amount of recognition element may affect the FRET efficiency.

FRET based sensing system is very appealing because of its simpleness of building ratiometric fluorescent systems. Unlike those of single-signal sensors, the ratiometric sensors contain two different fluorophores and use the ratio of the two fluorescence intensities to quantitatively detect the analytes. They can eliminate most ambiguities in the detection process by self-calibration of two emission bands. The external factors, such as excitation source fluctuations and concentration, will not affect the ratio between the two fluorescence intensities [80, 81].

Conventionally, the FRET-based sensing systems have been designed in the form of small molecular dyads, which contain two fluorophores connected by a spacer through covalent links [82] or a system containing the FRET pair and the recognition element in a controlled environment [83].

## 5. FRET based Ion sensor

FRET based detection has become a powerful tool for quantitative measurements of various analytes such as $H^+$ [84], metal ions [85, 86] and glucose [87] in environmental, industrial, medical, and biological applications because of its sensitivity, specificity, and real time monitoring with fast response time [88]. Recently Dibyendu et al [89] reported a method for the sensing of ions by determining the concentration of corresponding salts ($KCl$, $NaCl$, $MgCl_2$, $CaCl_2$, $FeCl_3$, $FeSO_4$, $AlCl_3$) in water, based on FRET between two laser dyes Acriflavine and Rhodamine B. The principle of the proposed sensor is based on the change of FRET efficiency between the dyes in presence of different ions ($K^+$, $Na^+$, $Mg^{2+}$, $Ca^{2+}$, $Fe^{2+}$, $Fe^{3+}$, $Al^{3+}$). Nano-dimensional clay platelet laponite was used to enhance the efficiency of the sensor.

Transition metal ions play an important role in biology as nutritional microelements as well as important ligands in proteins and small molecules. Sensing transition metal ions in biological systems is very crucial. The monitoring of toxic metal ions in aquatic ecosystems is an important issue because these contaminants can have severe effects on human health and the environment [90]. Lead and mercury are two of the most toxic metallic pollutants; for example, lead can cause renal malfunction and inhibit brain development [91] and mercury can damage the brain, heart, and kidneys [92]. Mercury pollution is a global problem and the major source of human exposure stems from contaminated natural waters [93]. Mercury undergoes long-range transport in the environment among various media such as air, soil, and water by deposition from anthropogenic releases. The atmospheric oxidation of mercury vapor to water-soluble $Hg^{2+}$ ions and its subsequent metabolism by aquatic microbes produces methyl mercury, a potent neurotoxin linked to many cognitive and motion disorders [94]. Thus, obtaining new mercury detection methods that are cost effective, rapid, facile and applicable to the environmental and biological milieus is an important goal. Liu et al [95] reported FRET based ratiometric sensor for the detection of $Hg^{2+}$ ion. Silica nanoparticles were labeled with a hydrophobic fluorescent nitrobenzoxadiazolyl dye which acts as a FRET donor. Rhodamine was then covalently linked to the surface of the silica particles which acts as acceptor. Nanoparticles are then exposed to $Hg^{2+}$ in water. The detection limit of this is 100 nM (ca. 20 ppb). FRET based system with control over the location of both donor and acceptor and their separation distance within the nanoparticles has been developed for ratiometric sensing of $Hg^{2+}$ in water [96]. A novel calyx(4) arene derivative locked in the 1,3-alternate conformation bearing two pyrene and rhodamine fluorophores was synthesized as a selective sensor for the $Hg^{2+}$ ion [97]. The principle of sensing was based on FRET from pyrene excimer emissions to ring opened rhodamine absorption upon complexation of the $Hg^{2+}$ ion.

Chao et al reported FRET based ratiometric detection system for mercury ions in water with polymeric particles as scaffolds [98]. A flexible 8-hydroxyquinoline benzoate linked Bodipy-porphyrin dyad has been designed, synthesized which can be used for selectively sensing of $Fe^{2+}$ and $Hg^{2+}$ ions [99].

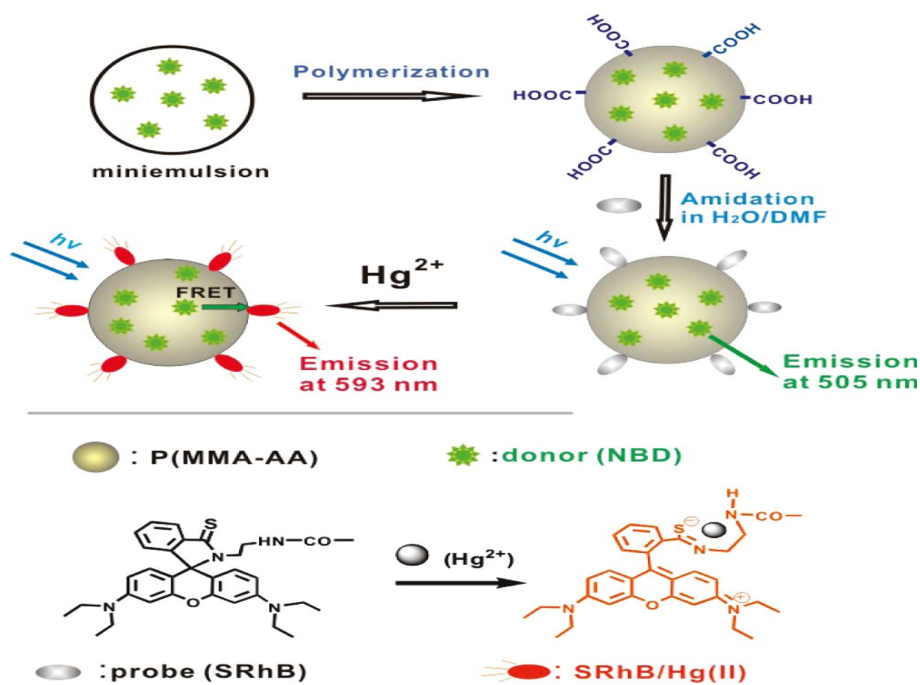

**Figure 2.** Formation of a FRET based system with polymeric nanoparticle as the Scaffold and its application as ratiometric fluorescence sensors for Mercury ion in water [98].

FRET-based ratiometric sensing platform based on β-cyclodextrin has been demonstrated [100]. β-cyclodextrin provides the hydrophilicity and biocompatibility; thus, the sensing platform can be used in aqueous medium and in some biological fluids as well as in live cells. Cyclodextrin based supramolecular complex has also been used for ratiometric sensing of ferric ion [101]. $Cr^{3+}$ ion is an essential trace element in human nutrition and has great impacts on the metabolism of carbohydrates, fats, proteins and nucleic acids by activating certain enzymes and stabilizing proteins and nucleic acids [102]. Based on the FRET from naphthalimide and rhodamine, $Cr^{3+}$-selective fluorescent probe for monitoring $Cr^{3+}$ in living cells with ratiometric fluorescent methods has been developed [103]

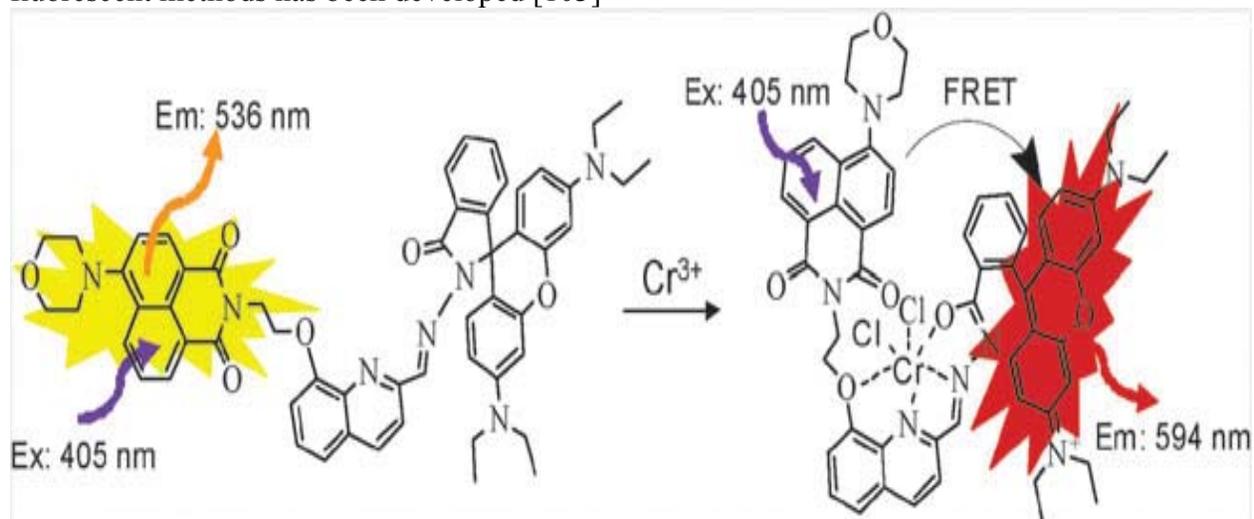

**Figure 3.** Mechanism of $Cr^{3+}$ -selective sensor [103]

A novel metal ion FRET sensor for potassium determination based on an ion-selective crown ether and energy transfer from carbon dots to graphene [104].

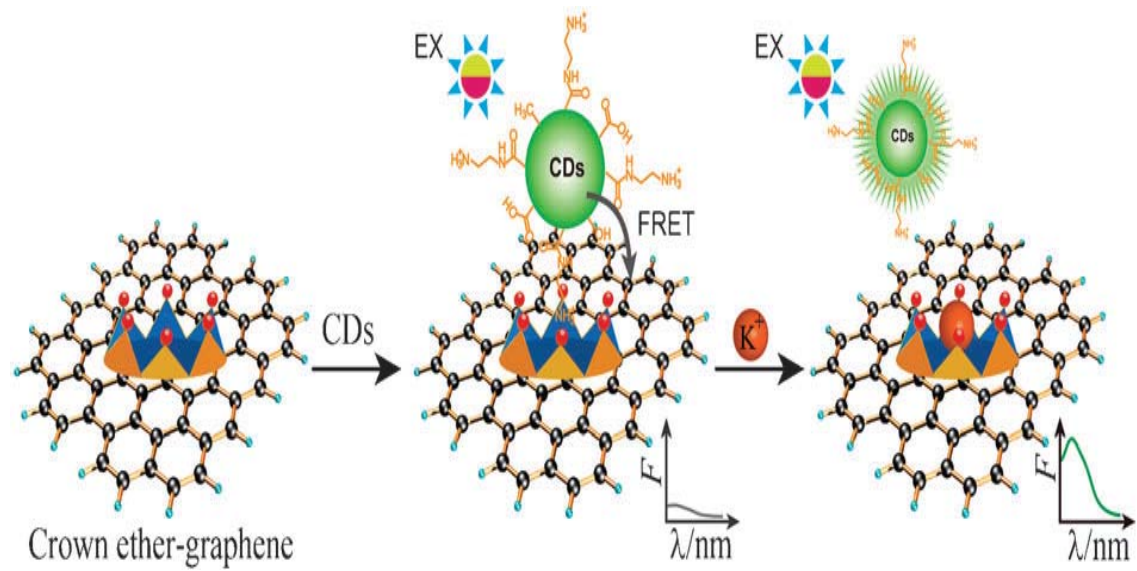

**Figure 4.** Schematic illustration of the FRET model based on carbon dots – graphene and the mechanism of $K^+$ determination [104].

Rhodamine based reversible chemosensor capable of undergoing excimer-fluorescent resonance energy transfer (Em-FRET) was designed to sense carboxylate anions using a ditopic receptor strategy [105]. Intramolecular FRET from the naphthalene emission to the coumarin absorption has been used to design ion sensor, which affords high fluorescence selectivity toward $F^-$ and $Cs^+$ ions [106]. Sensor containing guanidiniocarbonylpyrrole and a 9-(aminomethyl) anthracene moiety has been synthesized, which exhibits ratiometric fluorescence changes for $SO_3^{2-}$ over other anions. The change in fluorescence is attributed to the FRET and the $SO_3^{2-}$ complex induced photochemical reaction [107].

FRET based sensors have the potential to create time dependent concentration or activity maps of ions, small ligands, or macromolecules in living cells. In order to meet the challenge of multidimensional visualization, the dynamic range and response kinetics of the biosensors are critical attributes, since they directly affect the sensor's spatial and temporal resolution. Time-resolved microfluidic flow cytometer capable of characterizing the FRET based dynamic response of metal-ion sensors in mammalian cells has been designed [108]. The instrument can be used to examine the cellular heterogeneity of $Zn^{2+}$ and $Ca^{2+}$ sensor FRET response signals. Almost 30 fold difference between the extracellular and intracellular sensors has been reported [108]. FRET based $Cd^{2+}$ indicator containing a $Cd^{2+}$ binding protein obtained from Pseudomonas putida as the $Cd^{2+}$ sensing key has been reported capable of live cell dynamic sensing of $Cd^{2+}$ [109].

## 6. FRET based hard water sensor

The mineral content of "hard water" is very high in compared to "soft water". Though hard water is not harmful to one's health generally, but can cause serious problems in industrial settings, where water hardness should be monitored to avoid breakdowns of the costly equipments that handle water. The hardness of water is determined by the concentration of multivalent cations in water. The most common cations found in hard water include $Ca^{2+}$ and

$Mg^{2+}$. The temporary hardness in water involves the presence of dissolved carbonate minerals ($CaCO_3$ and $MgCO_3$), which can be reduced either by boiling the water or by addition of lime (calcium hydroxide) [110]. On the other hand the dissolved chloride minerals ($CaCl_2$ and $MgCl_2$) cause the permanent hardness of water that can not be removed easily, because it becomes more soluble as the temperature increases [111]. In that sense it is very important to identify the permanent hardness of water before use. Now a day the involvement of absorption or fluorescence spectroscopy for water analysis has received particular attention [112]. Sweetser and Bricker were the first who used the spectroscopic measurements to determine the concentration of calcium and magnesium ions in water [112]. Ion chromatography (IC) is another very powerful method for the analyses of anions and cations in aqueous solution [113]. Argüello and Fritz reported a method for the separation of $Ca^{2+}$ and $Mg^{2+}$ in hard water samples based on ion-chromatography and spectroscopic method [114]. E. Gömez et al reported a method for the simultaneous spectroscopic determination of calcium and magnesium using a diode-array detector [115]. The Fluorescence resonance energy transfer (FRET) phenomenon may be very effective tool for the designing of hard water sensors. Our group has designed a hard water sensor using FRET process [8]. To the best of our knowledge this could be the first attempt, where FRET process has been used for the detection of the hardness of water.

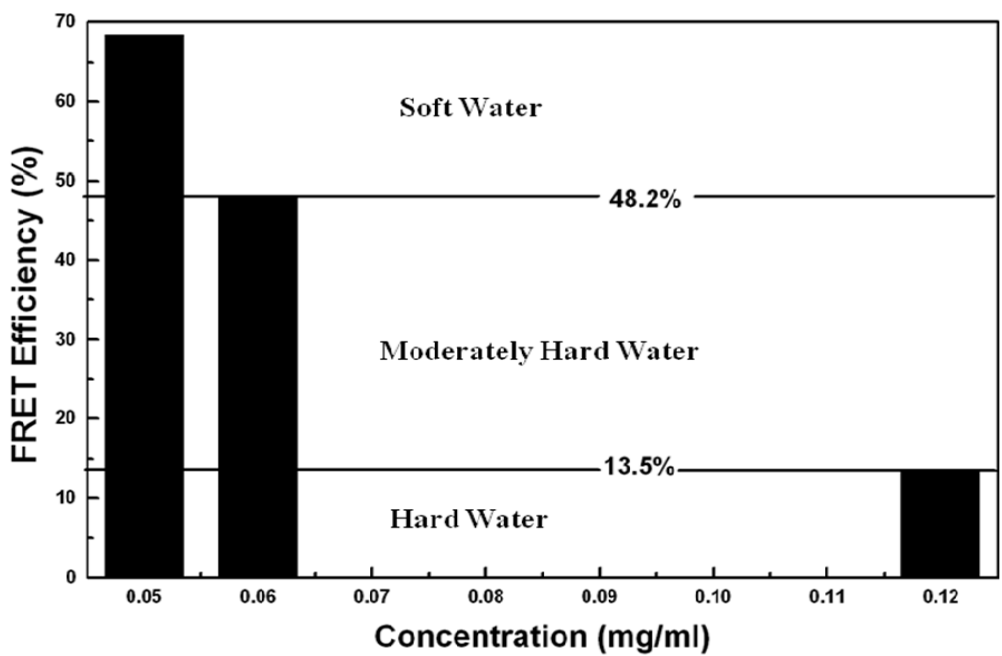

**Figure 5.** FRET efficiency of Acf and RhB mixture for the different concentration of $CaCl_2$ + $MgCl_2$ in presence of clay

In this case the effect of $Mg^{2+}$ or $Ca^{2+}$ or both on the FRET efficiency between two fluorophores, acraflavine (Acf) and rhodamine B (RhB) in presence of nanoclay sheet laponite has been investigated. The investigation showed that FRET efficiency decreases with increasing ion ($Mg^{2+}$ or $Ca^{2+}$ or both) concentration. This is because both the dyes Acf and RhB used were cationic in nature. Inclusion of cations increases the separation between them resulting a decrease in FRET efficiency. Nano clay platelet laponite was used to enhance the sensing efficiency. It has also been demonstrated that with proper calibration, this sensor can be used to sense water hardness with sufficient resolution between soft water (salt concentration less than 0.06 mg/ml), moderately hard water (salt concentration greater than 0.06 mg/ml and less than 0.12 mg/ml) and the hard water (salt concentration above 0.12 mg/ml) [110, 111].

## 7. FRET based pH sensors

The sensing of pH is one of the most powerful techniques which are essential in many fields of application ranging from agriculture and environment to industry, medicine and food. In medical science, abnormal pH values inside the cell indicate inappropriate cell function, growth and division. It is also helpful to diagnose some common disease like cancer and Alzheimer's. For the sensing of pH there are two very well known methods namely, (1) Optical chemical sensors (also termed optrodes) and (2) FRET based pH sensors. In case of optrodes the change in absorbance or fluorescence intensity of the pH sensitive dyes indicate a change in pH of the environment. On the other hand FRET based pH sensors are indicated by the ratiometric changes of the dye fluorescence of both donor and acceptor with pH of the environment.

Optrodes exploit pH indicator dyes (weak organic acids or bases) with distinct optical properties associated with their protonated (acidic) and deprotonated (basic) forms [116]. The absorption or fluorescence properties of these dyes are modified with a change of pH of the environment. A schematic representation of absorption- and emission-based sensing is shown in figure.

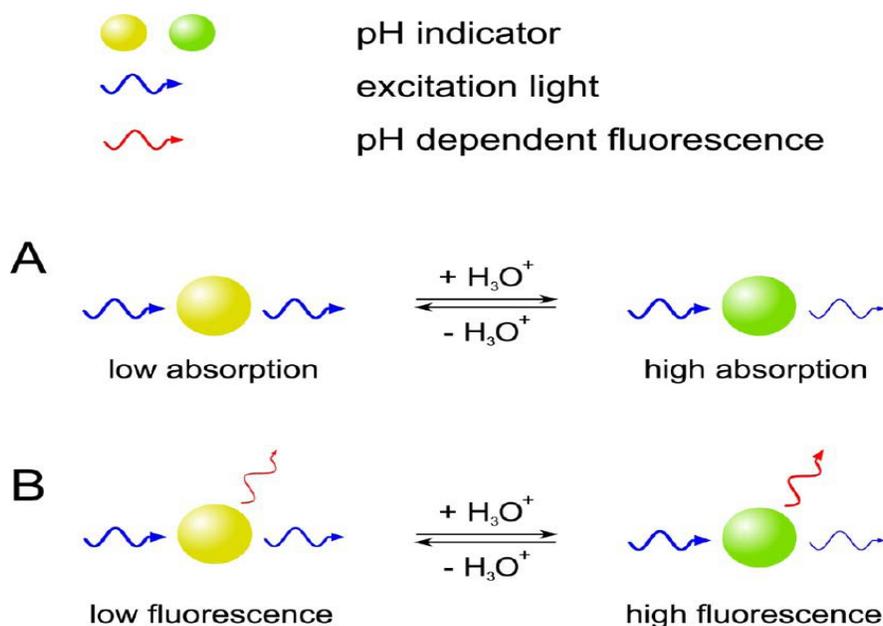

**Figure 6.** Schematic representation of the principle of (A) absorption based and (B) fluorescence-based pH sensing mechanisms [116].

There has been a consistent increase in output related to the development and application of optrodes since 1980. Apart from the overview sections, this review is exclusively focused on optical chemical pH sensors which have been developed in the last 2 years (2011−2013). Lin covers in his review different indicators, materials, and strategies for optical pH sensing published between 1991 and 2000. Fluorescent pH indicators offer better selectivity and sensitivity than absorption-based pH indicators. A review of fluorescent pH indicators has been published recently [117]. The most widely used fluorescent pH indicators are: 1-hydroxypyrene-3,6,8-trisulfonic acid (HPTS), fluorescein, and derivatives [118-120]. These dyes usually absorb in the visible blue region while emission occurs above 500 nm. pH sensors using the seminaphthofluorescein (SNAFL) have been also reported [121]. The fluorescein and SNAFL indicators exhibit poor photostability. HPTS exhibits excellent photostability, but its pH response is highly dependent on the ionic strength. pH-dependent lanthanide complexes have been also reported [122-125]. In 2007, novel pH sensitive coumarin-based indicators were described [126]. pH sensitive ruthenium metal−ligand complexes were also tested for use as luminescent pH indicators [127, 128]. Recently, Tormo et al. discussed the suitability of such complexes to be used as pH sensitive indicators [129].

Fluorescent sensors are one of the important methods for sensing of different chemical and biological materials but for this type of sensors change in fluorescent intensity could very well be perturbed by environmental factors. The introduction of FRET sensors can minimize this environmental perturbation, because it measures the ratio of two emissions in different environment. Energy transfer has been used for pH measurement [130]. Chan et al. demonstrated Förster resonance energy transfer (FRET)-based ratiometric pH nanoprobes where they used semiconducting polymer dots as a platform. The linear range for pH sensing of the fluorescein-coupled polymer dots was between pH 5.0 and 8.0 [131]. Egami et al. has introduced a fiber optic pH sensor, using polymer doped with either Congo red (pH range 3 to 5) or methyl red (pH range from 5 to 7) [132]. pH sensor based on the measurement of absorption of phenol red has also been reported [133], which can sense a pH range of 7 – 7.4. In one of our developed system of pH measurement using the change in FRET efficiency between Acf and RhB with pH we are capable of measuring over a wide range of pH 3.0 to 12.0 [9]. pH dependence of spectral overlap integral and FRET efficiency have been shown in figure 7. Sensing of wide range of pH using the present system is advantageous with respect to previous system [130].

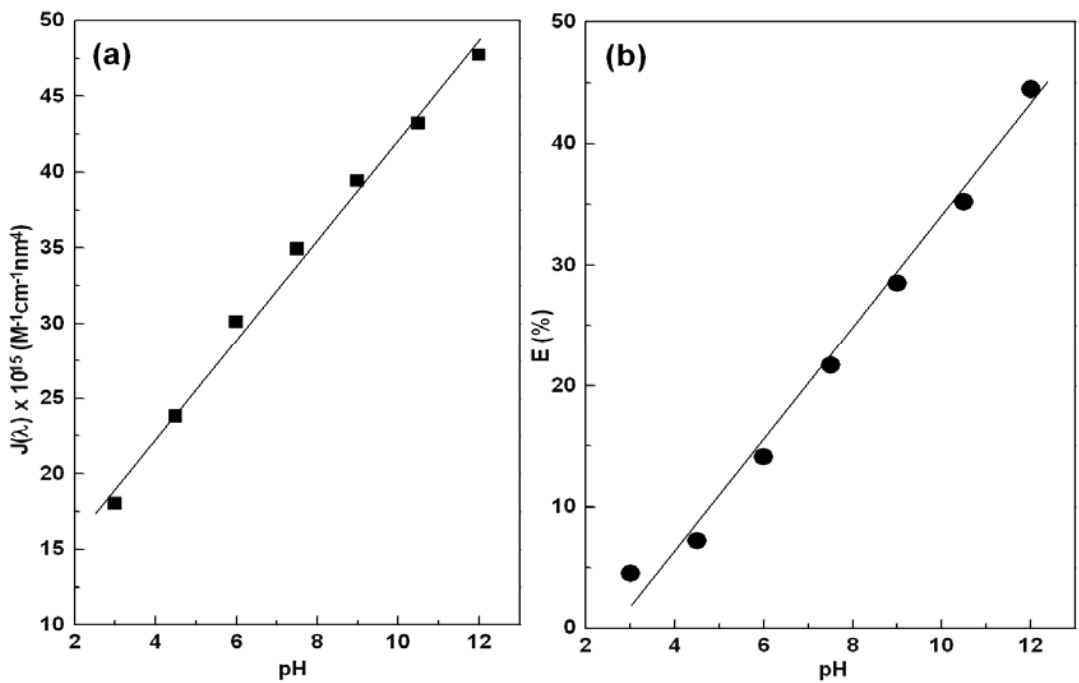

**Figure 7.** Variation of (a) spectral overlap integral ($J(\lambda)$) and (b) energy transfer efficiency (E) with increasing pH of the solution [9].

Intracellular pH is an important indicator for cellular metabolism and pathogenesis [131, 134]. pH sensing in living cells has been achieved using a number of synthetic organic dyes and genetically expressible sensor proteins, even allowing the specific targeting of intracellular organelles. Semiconducting polymer-based nanoparticles (Pdots) have recently emerged as a new class of ultrabright probes for biological detection and imaging. Poly (2,5-di(30,70-dimethyloctyl) phenylene-1,4-ethynylene) (PPE) Pdots has been used as a platform for designing FRET based ratiometric pH nanoprobes [131]. Where, a pH-sensitive dye, fluorescein was coupled to PPE Pdots offering a rapid and robust sensor for pH determination using the ratiometric methodology where excitation at a single wavelength results in two emission peaks, one that is pH sensitive and the other one that is pH insensitive for use as an internal reference. The linear range for pH sensing of the fluorescein-coupled Pdots is between pH 5.0 and 8.0, which is suitable for most cellular studies. The pH-sensitive Pdots show excellent reversibility and stability in pH measurements. This sensor has tested to measure the intracellular pH in HeLa cells following their uptake by endocytosis, thus demonstrating their utility for the use in cellular and imaging experiments [131]. pH lameleons are prototypes of a new class of pH sensors that can be further optimized, tuned, and targeted to different subcellular structures or attached to target proteins to interrogate pH changes in cellular microdomains [135]. pH lameleons possess ideal properties for intracellular pH measurements; they provide highly enhanced spectral dynamics compared to previous genetically encodable biosensors and offer fast and quantitative detection, with all the flexibility of genetically encodable biosensors. Esposito et al [135] reported a FRET-based pH sensor platform, based on the pH modulation of YFP acceptor fluorophores in a fusion construct with ECFP. Quantum dot-fluorescent protein FRET Probes for the sensing intracellular pH has been demonstrated [136] having high sensitivity and wide dynamic range, ratiometric measurements for internal calibration, dramatic reduction of photobleaching, and the ability to tailor the probe design for different pH ranges. These probes are well suited to a wide range of intracellular pH-dependent imaging applications that are not feasible with fluorescent proteins or organic fluorophores alone.

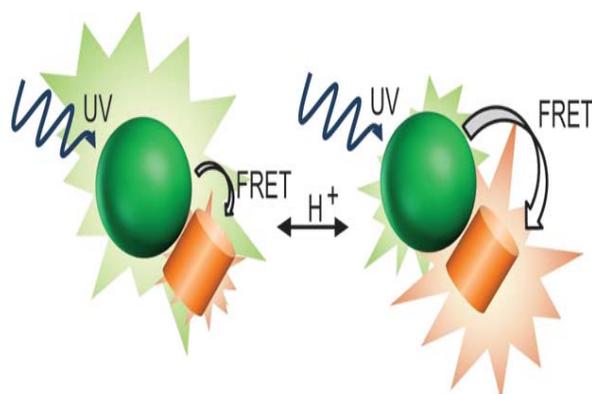

**Figure 8.** Schematic demonstration of the pH-dependent energy transfer between the quantum dot and fluorescent protein [121].

Semiconductor nanocrystals (NCs) serve as useful fluorescent labels owing to their photostability, continuous absorption spectra, and efficient, narrow, and tunable emission [137]. These properties of NCs have been exploited for applications in biological imaging and in single particle tracking studies [138]. A Ratiometric CdSe/ZnS Nanocrystal pH Sensor has also been reported [139, 140]. Taken together with the broad excitation spectrum and photostability conferred by NCs, the reversible and ratiometric approach presented here makes NCs versatile agents for chemical and biological sensing [139]. A new pH-sensitive polymeric sensor with dispersion stability and biocompatibility is synthesized, and its pH sensitivity is examined on the basis of the FRET efficiency [141].

## 8. FRET based Biosensors

Molecular activities in human body are highly dynamic and can occur locally in sub-cellular domains or compartments. Neighboring cells in the same tissue can exist in different states. Therefore, quantitative information on the cellular and sub-cellular dynamics of ions, signaling molecules, and metabolites is critical for functional understanding of organisms.
Biosensors are the devices that can measure enzyme activities, protein dynamics, and biophysical processes (e.g., membrane potential or molecular tension) or detect any disease related molecules using biological recognition element with sub-cellular resolution [142, 143]. Disease related molecules may be antibodies, antigens, nucleic acids or other biologically relevant small molecules, which are markers for a particular disease or condition [142, 143]. The biological recognition elements contain biomolecules such as antibodies, DNA probes or enzymes, which recognize a particular disease marker or analyte. The reaction between the biomolecule and the analyte results in physical or chemical changes that can result in the production of heat, mass, light, electrons or ions [144]. The analyte can also be labeled with a biomarker or tag, such as an enzyme, radioisotope or dye. If the dye is fluorescent, the device is then known as a fluorescence-based biosensor. The Pico-Quant GmbH system is an example of a very sensitive fluorescence based biosensor. It utilizes fluorescence decay lifetimes and it is possible to detect a single fluorescent molecule and its orientation with this system [145]. There is a demand for simple, compact, low-cost devices that can detect a low concentration of antigens. There are a few strategies being employed to enhance the sensitivity of fluorescence-based biosensors to achieve this goal. One of these strategies is FRET [146].

Quantitative and dynamic analysis of metabolites and signaling molecules is limited by technical challenges in obtaining temporally resolved information at the cellular and compartmental level. FRET technology enables the quantitative analysis of molecular dynamics in biophysics and in molecular biology, such as the monitoring of protein-protein interactions, protein-DNA interactions, and protein conformational changes. FRET-based biosensors have been utilized to monitor cellular dynamics not only in heterogeneous cellular populations, but also at the single-cell level in real time. Lately, applications of FRET-based biosensors range from basic biological to biomedical disciplines. Despite the diverse applications of FRET, FRET-based sensors still face many challenges. There is an increasing need for higher fluorescence resolution and improved specificity of FRET biosensors [147]. Additionally, as more FRET-based technologies extend to medical diagnostics, the affordability of FRET reagents becomes a significant concern.

In order to deduce the molecular mechanisms of biological function, it is necessary to monitor changes in the sub-cellular location, activation, and interaction of proteins within living cells in real time. Förster resonance energy-transfer (FRET)- based biosensors that incorporate genetically encoded, fluorescent proteins permit high spatial resolution imaging of protein−protein interactions or protein conformational dynamics. RhoA is a member of the Rho family, a subset of the Ras superfamily of GTP-binding proteins. RhoA is of central importance in the regulation of contractility and cell interaction with the extracellular matrix4. It serves as a molecular switch that cycle between a GDP-bound inactive state and a GTP-bound active state. Yousaf et al [148] reported a FRET-Based Biosensor to study the dynamics of RhoA GTPase activation in cells on patterned substrates.

The phenomenon of FRET between two fluorescent proteins of different hues provides a robust foundation for the design and construction of biosensors for the detection of intracellular events. Accordingly, FRET-based biosensors for a variety of biologically relevant ions, molecules, and specific enzymatic activities, have now been developed and used to investigate numerous questions in cell biology [142].

The most successful and highly exploited strategy to date in ratiometric multi-parameter fluorescence imaging has been to combine the use of a CFP/YFP FRET biosensor with a synthetic $Ca^{2+}$ indicator. Fortunately, several popular $Ca^{2+}$ indicators, including fura-2 and indo-1, either absorb or fluoresce at wavelengths that are distinct from that of the CFP/YFP pair [149].

Determination of infected or disease-related biomolecules in body fluids such as serum or plasma holds significant applications in clinical diagnosis. FRET based sensor has been designed and applied to monitor thrombin level in human plasma [150]. FRET from upconverting phosphors (UCP) to carbon nanoparticles (CNP) has been used to design this sensor. The sensor can be used for thrombin sensing both in an aqueous buffer and in a serum matrix with comparable performances, proving that the UCP-CNP FRET system is capable of overcoming background interference in complex biological samples [150].

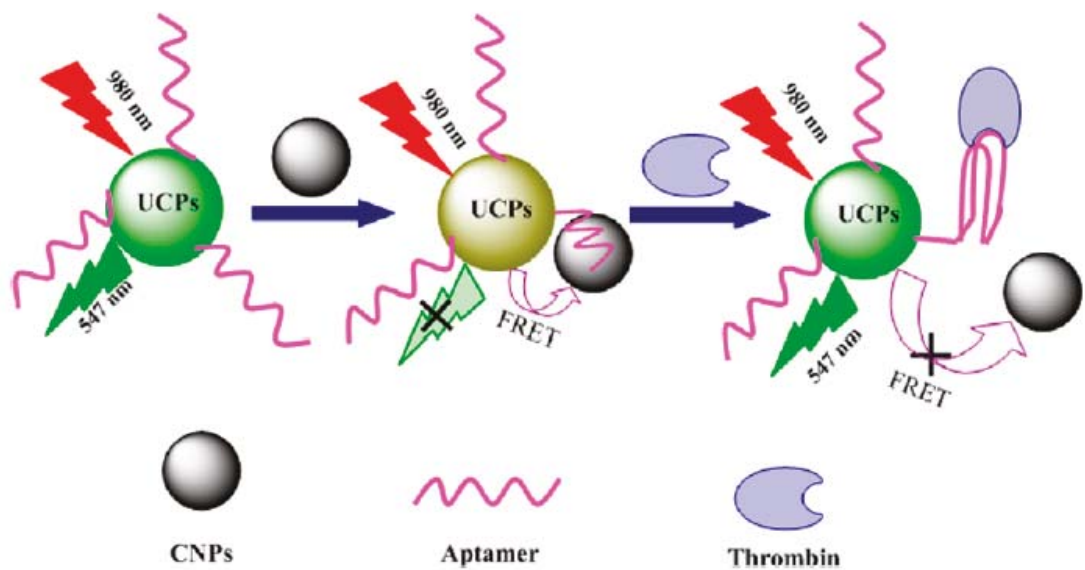

**Figure 9.** Schematic illustration of the Thrombin Sensor Based on FRET from Aptamer-Modified Upconverting Phosphors to Carbon Nanoparticles [150].

Matrix metalloproteinases (MMPs) are a family of zinc dependent endopeptides that degrades the extracellular matrix and basement membrane components. Specifically, MMP-2 (gelatinase A) is able to degrade type VI collagen and thus not only plays a key role in physiological and pathological states including morphogenesis, reproduction, and tissue remodeling, but also is one of the crucial MMPs in tumor growth, invasion, and metastasis [151, 152]. Due to the high complexity of the sample matrix, sensitive and selective determination of MMP-2 directly in blood samples is very difficult and a challenging job. Liu et al [153] has designed a new homogeneous biosensor for selective sensing of MMP-2 based on FRET from upconversion phosphors (UCPs) to carbon nanoparticles (CNPs). They also developed new upconversion FRET sensing platform using aromatic polymer nanospheres (poly-m-phenylenediamine, PMPD) as the energy acceptor [154], which may open the door for the new class of UC-FRET biosensors with wide applications [154].

Genetically-encoded FRET biosensors enable us to visualize a variety of signaling events, such as protein phosphorylation and G protein activation in living cells [155, 156]. Biosensors based on the principle of FRET have been developed to visualize the activities of the signaling molecules in living cells. Accordingly, FRET based biosensors have been used in cancer research [157]. Stable expression of FRET biosensors will accelerate current trends in cancer research, that is, from cells on a plastic dish to 3-D and/or live tissues, and from biochemistry to live imaging. A sensitive and specific FRET biosensor was developed by Mizutani et al [158] and applied to detect the activity of BCR-ABL kinase in live cells. This biosensor allowed the detection of cancerous and drug-resistant cells, and the evaluation of kinase inhibitor efficacy. This is an indication that future biosensor development and imaging using FRET can increasingly contribute to cancer diagnosis and therapeutics.

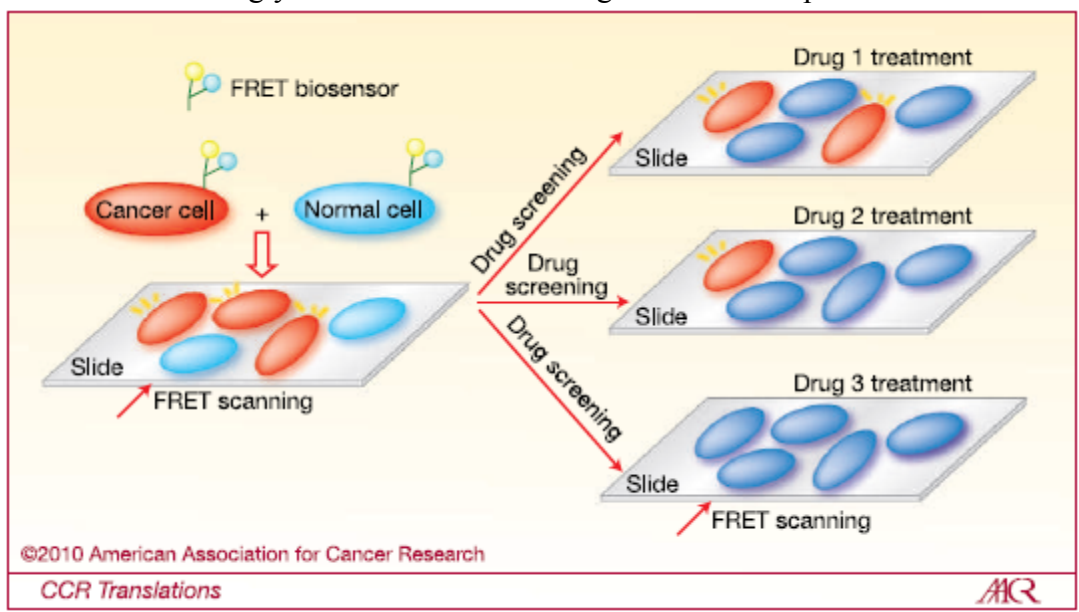

**Figure 10.** The application of a FRET biosensor for drug screening. Cancer (red) and normal (blue) cells from biopsy samples can be introduced with FRET biosensors to detect cancerous molecular activities, for example, BCR_ABL kinase activity. FRET scanning can identify the cancer cells and quantify their cancerous activities on the basis of the FRET signals. The biopsy samples and cells expressing the biosensors can be subjected to different drug treatments to assess the efficacy of different drugs in inhibiting the target molecular activities [158].

Simultaneous monitoring of multiple molecular interactions and multiplexed detection of several diagnostic biomarkers at very low concentrations have become important issues in advanced biological and chemical sensing. Optically multiplexed six-color FRET based biosensor for simultaneous monitoring of five different individual binding events has been reported [159]. Simultaneous FRET from one Tb complex to five different organic dyes measured in a filter-based time-resolved detection format with a sophisticated spectral crosstalk correction, which results in very efficient background suppression. The advantages and robustness of the multiplexed FRET sensor were exemplified by analyzing a 15-component lung cancer immunoassay involving 10 different antibodies and five different tumor markers in a single 50 μL human serum sample. Quantum-dot based FRET immunoassay for sensitive clinical diagnostics of low volume serum samples has also been demonstrated [160].

Schifferer et al [161] demonstrated a genetically encoded dynamic RNA reporter using intramolecular FRET between mutants of GFP. This may be useful in several types of application, for example, as reporter in vitro for real time studies on transcription or stability of RNA, to image very dynamic aspects of gene expression in vivo or to study relationships between RNA levels and protein expression in single living cells. FRET based biosensor has also been used for monitoring the σ1 receptor activation switch in living cells [162].

The use of conjugated polymers as biosensor devices is a growing research field, and the detection of small quantities of biomolecules is of great interest. Areas in which detection of DNA is of interest are for example forensic science, medical diagnostics, and the study of mutations. Indeed, the genomic revolution creates a great need for cheap methods for DNA detection and decoding. One approach to DNA detection is to use the special properties of conjugated polymers, which can respond to external stimuli, such as biomolecules, with a change in fluorescence spectra. Detection of specific DNA sequences has important applications in clinical diagnosis, the food and drug industry, pathology, genetics, and environmental monitoring.

One of the important methods for detecting the DNA hybridization method is by measuring the fluorescence signals, where the dye molecules are intercalated in to a DNA double helix [163]. But some inherent limitations of this method include a lack of specificity for many particular duplex and no possibility to create multiplexed assays [163]. Another most important strategy for the detection of DNA hybridization involves FRET. There are many reports where the detection and characterization of DNA involves FRET process. K. Fujimoto et al reported the detection of target DNAs by excimer-monomer switching of Pyrene using the FRET process [164]. DNA based nanomachine was reported by H. Liu et al using the FRET phenomenon [165]. Also for encrypting messages on DNA strands, various methods have been accomplished [166].
Influence of DNA presence on FRET efficiency between two laser dyes Acf and RhB has been studied [10]. Two types of molecular logic gates, namely, NOT and YES/NOT gate have been designed based on the FRET between Acf and RhB. These two molecular logic gates have been found efficient to detect the presence of DNA in aqueous solution having concentration as low as 1 μg/ml [10].

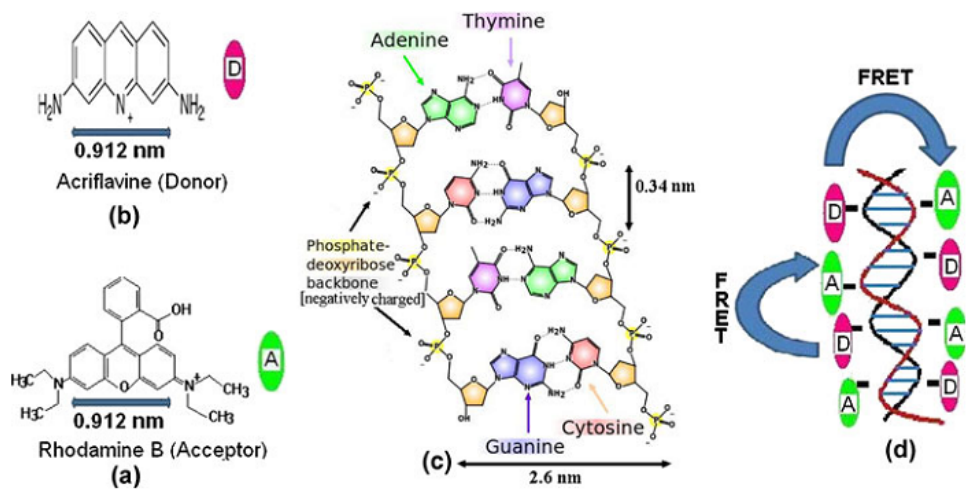

**Figure 11.** (a) Molecular structure of rhodamine B, (b) molecular structure of acriflavine, (c) structure of DNA showing the negatively charged phosphate deoxyribose backbone, (d) schematic diagram showing the attachment of Acf & RhB onto phosphate backbone of DNA [10].

The interactions between a zwitterionic polythiophene derivative, POWT, and DNA oligonucleotides in solution have been studied by FRET in order to demonstrate a DNA sensor [167]. A DNA biosensor based on FRET utilizing synthesized quantum dot (QD) has been developed for the detection of specific-sequence of DNA for Ganoderma boninense, an oil palm pathogen [168]. Detection of target DNA based on FRET has been demonstrated [169], where, introduction of graphene helped to reduce the background signal of traditional PFP-based DNA detection platform and thus enhances the sensing efficiency.

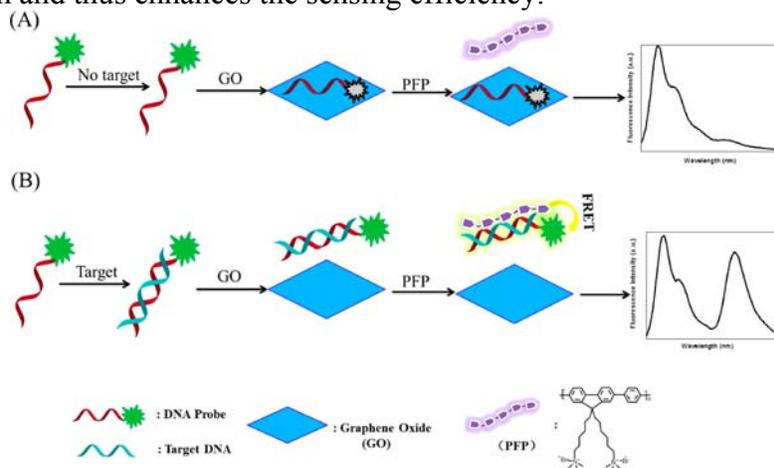

**Figure 12.** Schematic Representation of GO-Based Low Background-Signal Platform for the Detection of Target DNA [169].

Metal dependent global folding and activity of the 8-17 DNAzyme has been studied by FRET [170]. It has been shown that DNAzyme folds into compact structure(s) in the presence of $Zn^{2+}$ or $Mg^{2+}$ with stem III approaching a configuration constituting stems I and II without changing the angle between stems I and II. Activity, folding and Z-DNA formation of the 8-17 DNAzyme has also been studied in presence of monovalent ions [171].

Protein−DNA interactions play central roles in many biological processes. Studying sequence specific protein−DNA interactions and revealing sequence rules require sensitive and quantitative methodologies that are capable of capturing subtle affinity difference with high accuracy and in a high throughput manner. Double stranded DNA-conjugated gold nanoparticles and water-soluble conjugated polyelectrolytes are used as cooperative sensing elements to construct a suit of hybrid sensors for detecting protein−DNA interactions, exploiting the differential FRET with and without protein binding [172].

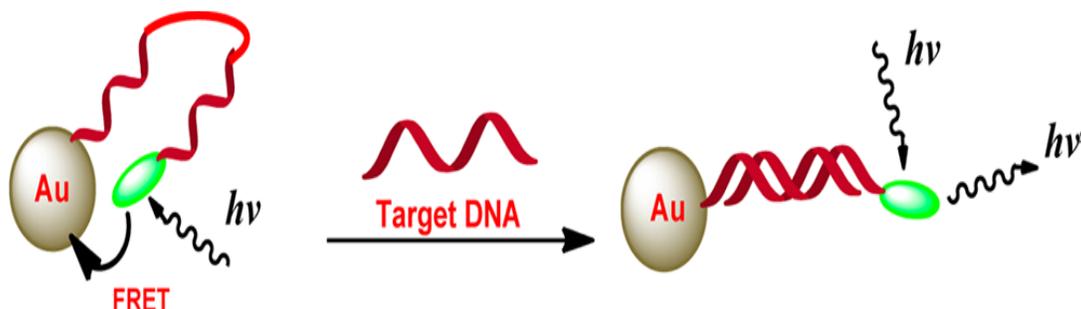

**Figure 13.** Schematic illustration of DNA detection, showing the conformational changes of dye−oligonucleotide−AuNP conjugates before and after hybridization with the target DNA [173].

Hairpin FRET-based systems for sensing DNA have been created by labeling molecular beacons with AuNPs. As shown in Figure 13, the nucleic acid sensor conjugated with the organic dye is self-complementary, forming the hairpin structure on AuNPs with effective FRET fluorescence quenching. The hairpin structure changes to rodlike through complementary hybridization with the target DNA, resulting in an increase in fluorescence of the dye. By employing similar principle, Nie et al. have shown that single stranded oligonucleotide-functionalized AuNPs with fluorophore-termini can assemble into a constrained arch-like conformation [174]. Mirkin et al. have developed AuNPs sensors, which are designed to detect and quantify intracellular analytes, for example, mRNA in cells [175].

The condensation and decondensation of DNA can be efficiently detected by FRET studies. Recently usefulness of intermolecular two-step FRET has been demonstrated [176], from quantum dots (QDs) on DNA to first a nucleic acid labeling dye and then Cy5 dye on the condensing agent, for the detection of DNA condensation [177]. Zhang et al have shown that water-soluble Cysteine-coated CdSe/ZnS QDs are capable of sensing the dissociation of DNA/polymer polyplexes [178]. The main advantage of these methods is that donors and acceptors labeled beyond the Förster distance in a DNA molecule precisely report the changes in the intramolecular conformation, the degree of condensation, and the stability of the condensed DNA [179, 180].

Phenomenon of FRET has been used to study DNA hybridization and cleavage processes. The hybridization was monitored by following FRET between Quantum Dots (QDs) and a molecular fluorophores (figure 14), [179, 181] whereas treatment of the QD/dye- DNA structure with deoxyribonuclease (DNase I) cleaved the DNA duplex and restored the fluorescence properties [182]. In general, FRET-based photodecomposition of excess acceptors in the proximity of highly photostable donors such as QDs will be helpful during the preparation of labeled DNA and other biomolecules for the analyses of not only DNA condensation and gene delivery but also protein-protein and protein-DNA interactions in biophysical investigations.

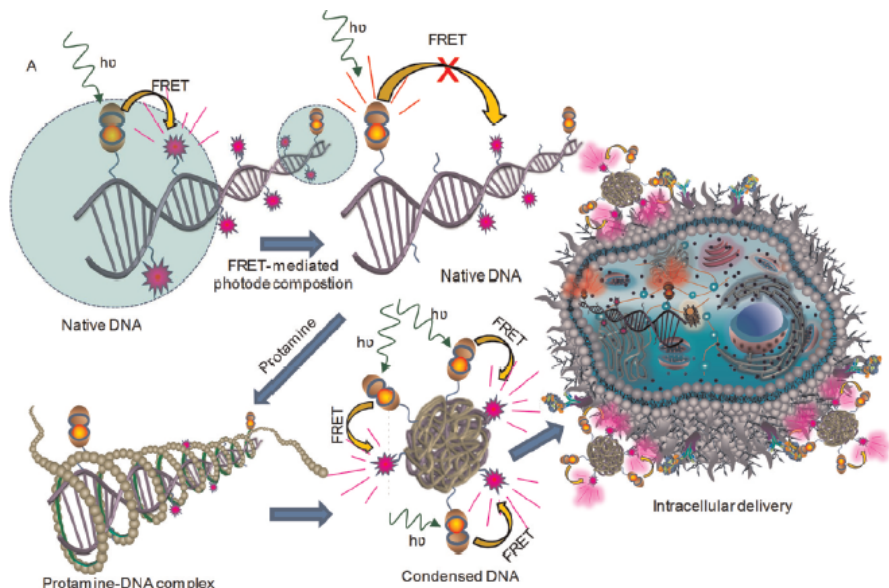

**Figure 14.** Hypothesis about FRET-mediated photodecomposition of undesired acceptors followed by the sensitive detection of the condensed DNA in a solution and decondensed DNA in a cell. The large number of acceptors left unaffected beyond the Förster distance can sensitively report the condensation of DNA by FRET-RAP [183].

Diabetes mellitus is a disease of major global concern, increasing in frequency at almost epidemic rates, such that the worldwide prevalence is predicted to at least double to about 300 million people over the next 10–15 years [184]. Diabetes is characterised by a chronically raised blood glucose concentration (hyperglycaemia), due to a relative or absolute lack of the pancreatic hormone, insulin. Therefore, control of glucose in blood is crucial for the long-term health of diabetics [185]. Continuous glucose monitoring (CGM) entered into clinical practice in 1999. Genetically encoded glucose nanosensors have been used to measure steady state glucose levels in mammalian cytosol, nuclei, and endoplasmic reticulum [186] Nevertheless, the performance of glucose sensors is widely thought to be a major bottleneck in the development of a closed-loop insulin delivery system (artificial pancreas) [187] where the clinical and regulatory requirements are for optimal accuracy and reliability of the CGM component.

Present methods to measure glucose require fresh blood, which is obtained by a finger stick. This procedure is painful and inconvenient, making it difficult to determine the glucose level in blood as frequently as is needed. Erratic blood glucose levels due to diabetes are responsible for adverse long-term problems of blindness and heart disease. These effects are thought to be due to glycosylation of protein in blood vessels. Consequently, there have been continued efforts to develop a noninvasive means to measure blood glucose and to develop fluorescence methods to detect glucose. These have often been based on the glucose-binding protein concanavalin A (ConA) and a polysaccharide, typically dextran, which serves as a competitive ligand for glucose (Figure 15a). Typically, the ConA is labeled with a donor (D) and the dextran with an acceptor (A), but the labels can be reversed. Binding of D-ConA to A-dextran results in a decrease in donor intensity or lifetime. The glucose in the sample competes for the glucose binding sites on D-ConA, releasing D-ConA from the acceptor. The intensity decay time and phase angles of the donor are thus expected to increase with increasing glucose concentration. This principle was used in the first reports of glucose sensing by fluorescence intensities. A fiber-optic glucose sensor was made using FITC-labeled dextran and rhodamine-labeled ConA (Figure 15b). The acceptor could be directly excited as a control measurement to determine the amount of Rh-ConA. The response of this glucose sensor is shown in Figure 15c. The donor and acceptor were placed on the dextran and ConA, respectively. The donor fluorescence was not completely recovered at high concentrations of glucose. This lack of complete reversibility is a problem that plagues ConA-based glucose sensors to the present day. It is expected that these problems can be solved using alternative glucose binding proteins, especially those that have a single glucose binding site and may be less prone to irreversible associations. It seems probable that site-directed mutagenesis will be used to modify the glucose-binding proteins to obtain the desired glucose affinity and specificity. As might be expected, lifetime-based sensing has been applied to glucose, and has been accomplished using nanosecond probes, long-lifetimes probes, and laser-diode-excitable probes. The problem of reversibility has been addressed by using sugar-labeled proteins in an attempt to minimize crosslinking and aggregation of the multivalent ConA. Such glucose sensors are occasionally fully reversible, but there is reluctance to depend on a system where reversibility is difficult to obtain.

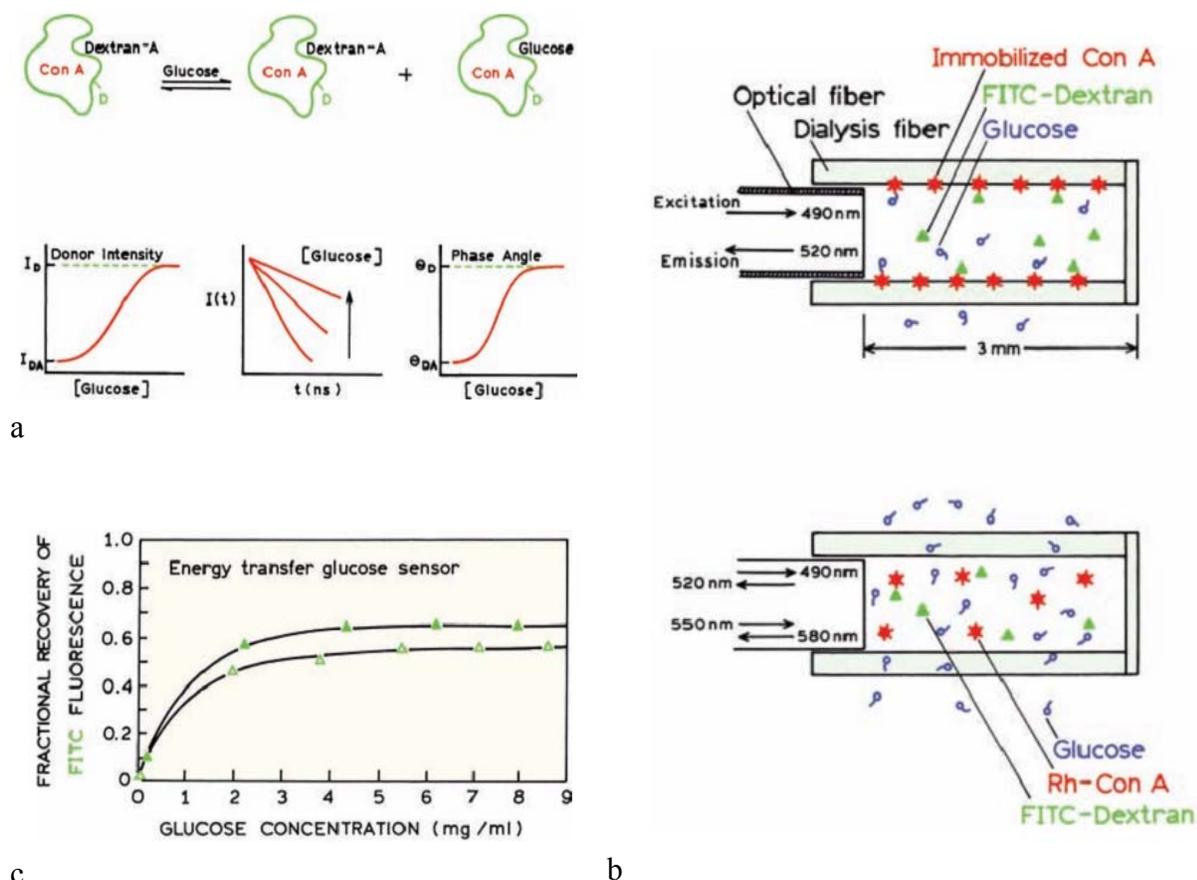

**Figure. 15.** Glucose sensing by resonance energy transfer. (a) Revised from [196], (b) Revised from [196], (c) Revised from [197].

Hsieh et al [188] demonstrated that multiple cysteines is a member of the periplasmic binding protein family can be selectively labeled with two different thiol-reactive reagents. This technique exploits a protein conformational change upon binding of a ligand, thus blocking one of the cysteine sites from the reaction chemistry. Using this technique for sequential labeling of glucose/galactose binding protein with the two dyes nitrobenzoxadiazole and Texas Red, two functional FRET sensors were prepared, and a glucose-dependent FRET signal was demonstrated for each of these. The ligand protection strategy may be of value for many further applications where dual-labeling of proteins is desired.

Dysfunction of endothelial cells under high glucose concentration is one of the major concerns for hyperglycemia [189]. A novel endothelial cell apoptosis detection method, which combines a caspase-3-based FRET biosensor cell line and the microfluidic chip system, was developed. The apoptosis and necrosis of endothelial cells undergo a series of different pulsatile flow conditions and glucose concentrations can be detected with high accuracy since the FRET biosensor provides real-time observation and does not need post treatment [190].

Precise and dynamic measurement of intracellular metabolite levels has been hampered by difficulties in differentiating between adsorbed and imported fractions and the sub-cellular distribution between cytosol, endomembrane compartments and mitochondria. Genetically encoded FRET based sensors were deployed for dynamic measurements of free cytosolic glucose and ATP at varying external supply and in glucose transport mutants [191]. These FRET sensors in a microfluidic platform, capable to monitor in vivo changes in intracellular free glucose in individual yeast cells. FRET based optical sensors for monitoring dynamic changes of intracellular metabolite levels in mammalian cells has also been reported [192]. Veetil et al [193] developed a new glucose sensor protein, AcGFP1-GBPcys-mCherry, and an optical sensor assembly, capable of generating quantifiable FRET signals for glucose monitoring [193]. This glucose sensor can generate measureable FRET signals in response to glucose concentrations varying from 25 to 800 μM. The possibility to measure glucose in vivo in the subconjunctival interstitial fluid for a period of 2 weeks was demonstrated in a human clinical trial [194]. It has been observed that a biocompatible surface coating on the implantable ocular mini implant enabled a longer duration of action of up to 6 months compared with 3 months for uncoated implants for in vivo glucose monitoring [195].

## 9. Conclusion and outlook

The present article has summarized recent developments of FRET based sensors with emphasis on biosensors. FRET is a unique phenomenon that combines the sensitivity and selectivity of fluorescence with the strong dependence of FRET on the distance between donor and acceptor molecules as well as their orientation. FRET is undoubtedly a powerful bioanalytical technique capable of making precise intramolecular measurements in a variety of experimental platforms and formats. Recent dramatic improvements in the development of fluorophores, such as fluorescent proteins and nanoparticles, along with the availability of advanced optical detection capabilities have enhanced the strength of this technique and resulted in its increasing popularity. Unlike with those of single signal sensors, the ratiometric FRET sensors eliminate most ambiguities in the detection by self-calibration of two emission bands of two different fluorophores. External factors, such as excitation source fluctuations and sensor concentration, will not affect the ratio between the two fluorescence intensities.

The demand for highly sensitive nonisotopic and noninvasive bioanalysis systems for biotechnology applications, such as those needed in clinical diagnostics, food quality control, and drug delivery, has driven research in the use of FRET for biological and chemical applications. Development of FRET based sensing system for practical application is a challenge, requiring an interdisciplinary outlook. Future progress of research in the area of FRET sensor is dependent upon the close collaboration of physicists, chemists, biologists, material scientists and computing specialists.

**Acknowledgements:**
The author SAH is grateful to DST and CSIR for financial support to carry out this research work through DST Fast-Track project Ref. No. SE/FTP/PS-54/2007, CSIR project Ref. 03(1146)/09/EMR-II.